\documentclass[prl,twocolumn,aps,showpacs,amsmath,amssymb]{revtex4}
\usepackage{graphics}

\begin{document}

\title{
  Computer aided perturbation theory by cumulants: dimerized and frustrated 
  spin 1/2 chain
}

\author{S. Sykora$^{a}$}
\author{A. H\"{u}bsch$^{a,b}$}
\author{K. W. Becker$^{a}$}
\affiliation{
  $^{a}$Institut f\"{u}r Theoretische Physik,
  Technische Universit\"{a}t Dresden, D-01062 Dresden, Germany \\
  $^{b}$Department of Physics, University of California, Davis, CA 95616
}

\date{\today}

\begin{abstract}
  This paper demonstrates that a computer aided perturbation theory can  
  easily be realized by use of a cumulant approach.
  In contrast to a recent alternative formulation on the basis of 
  Wegner's flow equation method the present approach can be applied to systems 
  with arbitrary Hilbert space. In particular an equidistant spectrum of the 
  unperturbed part of the Hamiltonian is not needed. The method is 
  illustrated in detail for dimerized and frustrated spin 1/2 chains for which 
  the ground state energy is calculated up to seventh order perturbation 
  theory. 
\end{abstract}


\maketitle


\section{Introduction}

Perturbation theory has proved very powerful for the investigation of problems 
which are not exactly solvable. Thus, a number of different schemes have been 
developed which are either based on canonical transformations (a well-known 
example is the Schrieffer-Wolff transformation \cite{Schrieffer}) or on 
projection technique \cite{Fulde}. Because of the increasing computer capacity 
one would like to perform such systematic perturbation expansions up to high 
orders by use of algebraic programmes. Recently developed 
computer aided algorithms are 
based on the flow equation method \cite{Wegner,Glazek,Stein} and Takahashi's 
formulation \cite{Takahashi} of standard perturbation theory. The latter
approach was used to derive an effective spin Hamiltonian for 
high-temperature superconductors \cite{M-Hartmann}, whereas the flow equation 
method has been applied to a number of low-dimensional systems, see 
e.g.~Refs.~\onlinecite{Knetter}. However, the applicability of the flow 
equation method is restricted to cases for which the unperturbed Hamiltonian 
has an equidistant eigenvalue spectrum. Only then the involved set of 
differential flow equations can be integrated.

Recently, a systematic perturbation expansion for many-particle 
systems in terms of cumulants has been proposed by two 
of us \cite{Hubsch}. This method is an projection approach and is based on 
the construction of effective Hamiltonians for low-energy properties.
Besides the automatically preserved size consistency of extensive variables, 
this cumulant method offers compact expressions for the different orders of 
the perturbation theory.  The main problem for the evaluation of 
cumulant expressions is to properly count the number of contributing 
configurations or 'diagrams' to each order. However, since the counting
of  configurations can be taken over by the computer, the cumulant approach is 
an ideal starting point for an algebraic computer aided evaluation of 
physical quantities. The aim of the present paper is to demonstrate how 
these ideas can be put into practice.

For the sake of comparability, let us  consider a dimerized and frustrated 
$S=1/2$ spin chain
\begin{eqnarray}
  \label{G1}
  {\cal H} &=& {\cal H}_{0} + {\cal H}_{1}\\
  {\cal H}_{0} &=& J \sum_{j=1}^{N}{\bf s}_{j}^{e}\cdot{\bf s}_{j}^{o} 
  \nonumber \\
  {\cal H}_{1} &=& J \lambda \sum_{j=2}^{N}
  \left\{
    {\bf s}_{j}^{e}\cdot{\bf s}_{j-1}^{o}+
    \alpha
    \left(
      {\bf s}_{j}^{e}\cdot{\bf s}_{j-1}^{e} + 
      {\bf s}_{j}^{o}\cdot{\bf s}_{j-1}^{o}
    \right)
  \right\}
  \nonumber
\end{eqnarray} 
in the limit of strong dimerization. This model was recently also 
investigated in Ref.~\onlinecite{Knetter} by use of the flow equation method.   
In Eq.~\eqref{G1}, ${\bf s}_{j}^{e}$ (${\bf s}_{j}^{o}$) denotes the spin 
at even (odd) site of dimer $j$. The unperturbed part ${\cal H}_{0}$ of the 
Hamiltonian \eqref{G1} describes $N$ uncoupled dimers. Therefore, its ground 
state is the product of singlets on all dimers, and the excited states of 
${\cal H}_0$ can be classified with respect to the number of 
local triplets. The perturbation ${\cal H}_{1}$ describes exchange
interactions between neighboring dimers. In literature, spin chains of 
type \eqref{G1} have been used for some spin-Peierls compounds like 
CuGeO$_{3}$ or TTFCuBDT \cite{Riera,Castilla,Bray}. Note that 
\eqref{G1} is an example for a model with an equidistant unperturbed level 
scheme.

The paper is organized as follows. In the next section the cumulant 
approach \cite{Hubsch} is formulated. In particular, the 
perturbation expansion of the resulting effective Hamiltonian is given 
in terms of cumulant expressions. In Sec.~\ref{chain} we develop 
the computer aided perturbation theory based on cumulants for the 
dimerized and frustrated $S=1/2$ chain.  
The ground state energy is calculated up to seventh order in the 
interaction. This model is generalized in Sec. \ref{general} so that the 
spectrum of the unperturbed part ${\cal H}_{0}$ of the Hamiltonian becomes 
non-equidistant. Finally, the conclusions are presented in Sec.~\ref{Conc}.


\section{Cumulant approach} \label{Cumulant}

The cumulant approach \cite{Hubsch} starts from the decomposition of the 
Hamiltonian ${\cal H}={\cal H}_{0}+{\cal H}_{1}$ into an unperturbed part 
${\cal H}_{0}$ and into a perturbation ${\cal H}_{1}$. The Hilbert space of 
the unperturbed Hamiltonian ${\cal H}_{0}$ is split into two subspaces: The 
low-energy part ${\cal U}_{P}$ and the high-energy part ${\cal U}_{Q}$ with 
projection operators ${\cal P}$ and 
${\cal Q}={\bf 1}-{\cal P}$. Let us assume that the two 
subspaces are separated by a finite energy difference. It is our 
aim to construct an effective Hamiltonian for the low-energy 
subspace ${\cal U}_{P}$.

Motivated by the quantum statistical expression for the free 
energy, the effective Hamiltonian for the subspace 
${\cal U}_{P}$ is defined as follows
\begin{eqnarray}
  \label{G2}
  {\cal H}_{\rm eff} &=& 
  -\frac{1}{\beta}{\cal P} \ln 
  \left(
    e^{-\beta{\cal H}}
  \right)_{P} {\cal P}
\end{eqnarray}
where $(\cdots)_P$ denotes the operator product 
${\cal P}(\cdots){\cal P}$, $\beta$ is the inverse 
temperature \cite{Polatsek,Hubsch}. Note that due to the projectors 
${\cal P}$ in Eq. \eqref{G2} the effective Hamiltonian ${\cal H}_{\rm eff}$ 
only acts in the low-energy subspace ${\cal U}_{P}$.

In order to transform Eq.~\eqref{G2} into a cumulant expression we introduce 
generalized cumulants
\begin{eqnarray}
  \label{G3}
  \lefteqn{
    \left(
      {\cal X}_{1}^{\nu_{1}} \cdots {\cal X}_{N}^{\nu_{N}}
    \right)_{P}^{C}
    \, \stackrel{\mathrm{def}}{=} \,
  } 
  && \\
  &=&
  \left.
    \frac{\partial^{\nu_1}}{\partial\xi_{1}^{\nu_{1}}} \cdots
    \frac{\partial^{\nu_N}}{\partial\xi_{N}^{\nu_{N}}} 
    \ln 
    \left( 
      e^{\xi_{1}{\cal X}_{1}} \cdots e^{\xi_{N}{\cal X}_{N}}
    \right)_{P}
  \right|_{\xi_{i}={0}\mbox{ } \forall{i}}
  \nonumber
\end{eqnarray} 
which in contrast to usual cumulants \cite{Kubo,Kladko} are still operator 
quantities. A detailed discussion of generalized cumulants 
can be found in Ref.~\onlinecite{Hubsch}. By use of series expansions we  
transform the effective Hamiltonian \eqref{G2} into a compact cumulant 
expression. Its Laplace transform can be used to derive a perturbation 
series for the effective Hamiltonian. For the case that  
all states of the relevant ${\cal U}_{P}$ subspace are degenerate 
with respect to ${\cal H}_{0}$, the
resulting effective Hamiltonian reads at temperature $T=0$ \cite{Hubsch}
\begin{eqnarray}  
  \label{G4}
  \lefteqn{
    {\cal H}_{\rm eff} \left( \beta\rightarrow\infty \right)
    \, = \,
  }
  && \\
  &=& 
  \left( {\cal H}_{0} \right)_{P}^{C}
  + \lim_{z\to0}
  \left\{
    \sum_{n=0}^{\infty}
    \left(
      {\cal H}_{1}
      \left[
        \frac{1}{z - {\bf L}_{0}}{\cal H}_{1}
      \right]^{n}
    \right)_{P}^{C}
  \right\}.
  \nonumber
\end{eqnarray}
Here, ${\bf L}_{0}$ is the Liouville operator with respect to ${\cal H}_{0}$. 
It is defined by ${\bf L}_{0} {\cal A} = [{\cal H}_{0}, {\cal A}]$ for any 
operators ${\cal A}$.

To calculate the cumulants in Eq.~\eqref{G4} one first 
decomposes the perturbation 
${\cal H}_{1}$ into eigenoperators of ${\bf L}_{0}$
\begin{eqnarray}
  \label{G5}
  {\cal H}_{1} &=& \sum_{m} {\cal T}_{m} 
  \qquad\mbox{with} \quad
  {\bf L}_{0}{\cal T}_{m} \,=\, \Delta_{m}{\cal T}_{m}. 
\end{eqnarray}
Due \eqref{G5} also products of 
${\cal T}_{m}$ are  eigenoperators of ${\bf L}_{0}$ \cite{Polatsek}. 
Therefore, the energy denominators of 
Eq.~\eqref{G4} can directly be evaluated
\begin{eqnarray}
  \label{G6}
  \lefteqn{
    \left(
      {\cal A} \frac{1}{ z - {\bf L}_{0} }
      {\cal T}_{1} \cdots {\cal T}_{M}
    \right)_{P}^{C}
    \, = \,
  }
  && \\
  &=&
  \frac{1}{ z - (\Delta_{1} + \cdots + \Delta_{M}) }
  \left(
    {\cal A} {\cal T}_{1} \cdots {\cal T}_{M}
  \right)_{P}^{C} .
  \nonumber
\end{eqnarray}
We are left with the calculation of cumulant expressions of the general form 
$({\cal T}_{1} \cdots {\cal T}_{M})_{P}^{C}$. For this purpose, we rewrite 
the cumulant expression \eqref{G3}  by expanding the 
logarithm into powers of $\xi_{i}$ and perform the 
differentiations.  We find the following decomposition of the generalized
cumulants into operator products \cite{Hubsch}
\begin{widetext}
\begin{eqnarray}
  \label{G7}
  ({\cal T}_{1} \cdots {\cal T}_{M})_{P}^{C}
  &=&
  ({\cal T}_{1} \cdots {\cal T}_{M})_{P}
  - \frac{1}{2}
  \sum_{
    \genfrac{}{}{0pt}{1}{n_{1},\dots,n_{M}=0}
    {(n_{1},\dots,n_{M})\not=(0,\dots,0)}
  }^{1}
  \sum_{
    \genfrac{}{}{0pt}{1}{m_{1},\dots,m_{M}=0}
    {(m_{1},\dots,m_{M})\not=(0,\dots,0)}
  }^{1}
  \delta(1, n_{1} + m_{1}) \cdots \delta(1, n_{M} + m_{M}) \times \\
  &&
  \qquad\qquad\qquad\qquad\qquad\qquad\qquad\qquad\qquad
  \times 
  ({\cal T}_{1}^{n_{1}} \cdots {\cal T}_{M}^{n_{M}})_{P}
  ({\cal T}_{1}^{m_{1}} \cdots {\cal T}_{M}^{m_{M}})_{P}
  + \cdots .\nonumber
\end{eqnarray}
\end{widetext}
Thus, the calculation of cumulant expressions is reduced 
to the evaluation of sums over operator products which can be 
easily done by use of a computer. The main limitation 
for a concrete realization is given by the increasing 
number of convoluted sums. Consequently, the numerical effort 
may be considerable,
if additional restrictions have to be taken into account.


\section{Dimerized and frustrated spin 1/2 chain} \label{chain}

In this section we show how the cumulant approach 
can be used to perform specific calculations. For this purpose we want to 
construct an effective Hamiltonian for the dimerized and frustrated spin 1/2 
chain \eqref{G1} in the limit of strong dimerization. As mentioned above, in 
this limit the unperturbed part ${\cal H}_{0}$ of the Hamiltonian describes 
isolated dimers without interaction between different dimers. Therefore, 
the low-energy subspace ${\cal U}_{P}$ is given by a single state which is a 
product state formed by singlets on all dimers. 
The low- and the high-energy 
subspaces are separated by the singlet-triplet splitting
\begin{eqnarray}
  \label{G8}
  \Delta &=& \varepsilon_{t} - \varepsilon_{s} = J
\end{eqnarray}
on a single dimer. Thus, the low-energy subspace ${\cal U}_{P}$ only 
consists of a single state, i.e. the singlet product state.  
The effective Hamiltonian acting in 
${\cal U}_{P}$ can be directly identified with the 
ground-state energy of the complete problem, multiplied by
the projector ${\cal P}$. 

\begin{table}
  \begin{center}
    \scalebox{0.61}{
      \includegraphics*[120,490][510,690]{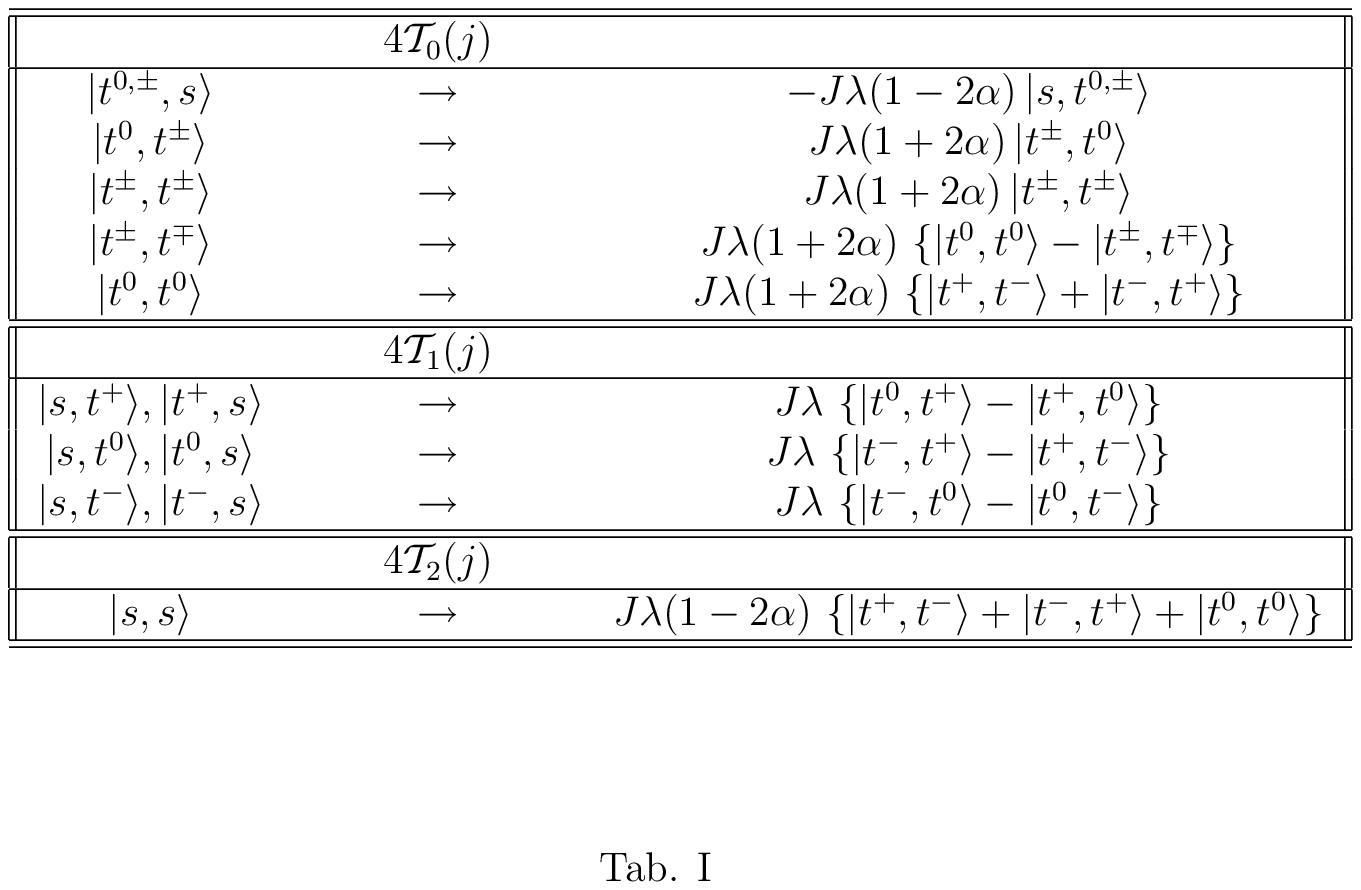}
    }
  \end{center}
  \caption{
    Action of the ${\cal T}_{m}(j)$ as used in the calculations. For 
    convenience, the dimer indices of the states are suppressed.
  }\label{eigen}
\end{table}

Next we decompose the perturbation ${\cal H}_{1}$ into eigenoperators 
of the Liouville operator ${\bf L}_{0}$. 
As mentioned above, the excited states of 
the unperturbed Hamiltonian ${\cal H}_{0}$ [see Eq.~\eqref{G1}] can be 
classified according to the number of local triplets. The 
creation and the annihilation of a local triplet state can be 
interpreted as the fundamental excitation processes. Therefore, also
the eigenoperators of the 
Liouville operator can be classified by the number of local 
triplets and ${\cal H}_{1}$ can be rewritten as
\begin{eqnarray}
  \label{G9}
  {\cal H}_{1} &=& 
  {\cal T}_{-2} + {\cal T}_{-1} + {\cal T}_{0} + {\cal T}_{1} + {\cal T}_{2}.
\end{eqnarray} 
An eigenoperator ${\cal T}_{m}$ creates $m$ local triplets. The 
respective eigenvalues of the Liouville operator are 
\begin{eqnarray}
  \label{G10}
  \Delta_{-2} &=&   -2 \Delta, \qquad
  \Delta_{-1} \,=\, -  \Delta, \qquad
  \Delta_{0}  \,=\,  0       , \\
  \Delta_{1}  &=&    \Delta, \qquad
  \Delta_{2}  \,=\,  2 \Delta. \nonumber
\end{eqnarray}
The perturbation ${\cal H}_{1}$ [see Eq.~\eqref{G1}] consists of 
interactions between adjacent dimers. Therefore, 
no further excitation operators 
occur, and the ${\cal T}_{m}$'s can be directly constructed from 
two-dimer matrix elements
\begin{widetext}
\begin{eqnarray}
  \label{G11}
  \langle x_{j-1} , x_{j} |
  J \lambda \left\{
    {\bf s}_{j}^{e}\cdot{\bf s}_{j-1}^{o}+
    \alpha
    \left(
      {\bf s}_{j}^{e}\cdot{\bf s}_{j-1}^{e} + 
      {\bf s}_{j}^{o}\cdot{\bf s}_{j-1}^{o}
    \right)
  \right\}
  | y_{j-1} , y_{j} \rangle
\end{eqnarray}
\end{widetext}
where $x_{j-1}$, $y_{j-1}$, $x_{j}$, $y_{j}$ denote singlets ($s$) and 
triplets ($t^{-}$, $t^{0}$, $t^{+}$) on the dimers $(j-1)$ and $j$. 
Let us introduce local transition operators ${\cal T}_m(j)$
by the relation 
\begin{eqnarray}
  \label{G12}
  \lefteqn{
    {\cal T}_{-2}(j) + {\cal T}_{-1}(j) + {\cal T}_{0}(j) + 
    {\cal T}_{1}(j) + {\cal T}_{2}(j)
    \, = \,
  }
  \qquad && \\
  &=&
  J \lambda \left\{
    {\bf s}_{j}^{e}\cdot{\bf s}_{j-1}^{o}+
    \alpha
    \left(
      {\bf s}_{j}^{e}\cdot{\bf s}_{j-1}^{e} + 
      {\bf s}_{j}^{o}\cdot{\bf s}_{j-1}^{o}
    \right)
  \right\}\nonumber
\end{eqnarray}
which are again classified with respect to the net 
change of the number of triplets. 
Eqs.~\eqref{G12} together with  \eqref{G11} allows to construct
the transition operators.  
For instance, matrix elements of \eqref{G11}
 connecting two singulet states [on the dimers $(j-1)$ 
and $j$] with two triplet states  
[on the dimers $(j-1)$ and $j$] contribute to 
${\cal T}_{2}(j)$. The results for ${\cal T}_0(j), {\cal T}_1(j)$,
${\cal T}_2(j)$ are summarized in Tab. \ref{eigen}. 
The remaining transition operators ${\cal T}_{-1}(j)$ 
and ${\cal T}_{-2}(j)$ follow from the relation 
${\cal T}_{-m}(j) = {\cal T}_{m}^{\dagger}(j)$. The 
compact eigenoperators ${\cal T}_m$ of 
the Liouville operator are given by
\begin{eqnarray}
  \label{G13}
  {\cal T}_{m} &=& \sum_{j=2}^{N} {\cal T}_{m}(j).
\end{eqnarray}
Note that the above decomposition of ${\cal H}_{1}$ was also derived 
in Ref.~\onlinecite{Knetter}.  

Now we can calculate the cumulant expressions
\begin{eqnarray}
\label{G14}
  \lefteqn{
    \left(
      {\cal T}_{m_{1}} \dots {\cal T}_{m_{k}}
    \right)_{P}^{C}
    \, = \,
  }
  && \\
  &=&
  \sum_{j_{1}=2}^{N} \dots \sum_{j_{k}=2}^{N}
  \left(
    {\cal T}_{m_{1}}(j_{1}) \dots {\cal T}_{m_{k}}(j_{k})
  \right)_{P}^{C}
  \nonumber
\end{eqnarray}
which enter the effective Hamiltonian \eqref{G4}. For that purpose 
it is favorable to 
exploit some additional restrictive conditions.  In this way the 
number of contributions to the cumulants, which have to be calculated 
explicitly,  can be considerably reduced. Remember, the effective 
Hamiltonian \eqref{G4} acts in the low-energy subspace 
${\cal U}_{P}$ which is formed by the product state of singulets on all
dimers. It follows   
\begin{itemize}
  \item[(i)]
  Products of the ${\cal T}_{m}$ operators inside the cumulant expression 
  do not change the number of triplets, i.e. 
  $m_{1} + m_{2} + \dots + m_{k}=0$ has to be fulfilled.
  \item[(ii)]
  Due to the explicit form of the excitation operators 
(see Tab.~\ref{eigen}), the low-energy
  subspace ${\cal U}_{P}$ and the high-energy subspace ${\cal U}_{Q}$ are only
  connected by ${\cal T}_{2}$ and ${\cal T}_{-2}$. Thus, 
  non-vanishing contributions to (\ref{G14}) can only occur 
if $m_{1}=-2$ and $m_{k}=2$.
  \item[(iii)]
  Due to general properties of cumulants \cite{Kubo} only connected processes 
  or 'diagrams' contribute to cumulant expressions. Therefore, the maximum 
  range of connected dimers in \eqref{G14} is restricted to 
  \begin{eqnarray}
    \label{G15}
    \lefteqn{
      j_{\rm max} - j_{\rm min} 
      \, = \,
    }
    && \\
    &=& 
    {\rm max(j_{1},\dots,j_{k})} - {\rm min}(j_{1},\dots,j_{k}) \,\le\, (k-1)
    \nonumber
  \end{eqnarray}
  All dimers between $j_{\rm min}$ and $j_{\rm max}$ enter the 
  cumulant.
\end{itemize}
Whereas (i) and (ii) directly follow from Eqs.~\eqref{G14} and \eqref{G7} 
the condition (iii) represents a basic feature of cumulant expressions. Note 
that these restrictive conditions extremely simplify the evaluation of the 
effective Hamiltonian \eqref{G4}. The zeroth order of ${\cal H}_{\rm eff}$ is
given by ${\cal H}_{0} = N\varepsilon_{s}{\cal P}$ whereas the first order
contribution vanishes due to $({\cal H}_{1})_{P}=0$. The second order 
contribution to the effective Hamiltonian reads 
\begin{eqnarray}
  \label{G16}
  {\cal H}_{\rm eff}(\beta\rightarrow\infty)|_{\rm 2nd\, order} &=&
  - \left(
    {\cal H}_{1} \frac{1}{{\bf L}_{0}} {\cal H}_{1}
  \right)_{P}^{C} \\
  &=&
  - \frac{1}{2\Delta}
  \left(
    {\cal T}_{-2} {\cal T}_{2}
  \right)_{P}^{C} 
  \nonumber
\end{eqnarray}
where we have used (ii). Now we insert Eq. \eqref{G13} into \eqref{G16} and 
take the condition (iii) into account. Thus
\begin{eqnarray}
  \label{G17}
  \lefteqn{
    {\cal H}_{\rm eff}(\beta\rightarrow\infty)|_{\rm 2nd\, order}
    \, = \,
  }
  && \\
  &=&
  - \frac{1}{2\Delta}
  \sum_{j_{1}=2}^{N}
  \sum_{j_{2}=2}^{N}
  \left(
    {\cal T}_{-2}(j_{1}) \, {\cal T}_{2}(j_2)
  \right)_{P}^{C} \nonumber \\
  &=&
  - \frac{N}{2\Delta}
    \left(
    {\cal T}_{-2}(j) \, {\cal T}_{2}(j)
  \right)_{P}^{C}
  \nonumber 
\end{eqnarray}
where the dimer site index 
$j$ can be chosen arbitrarily. Note that in \eqref{G17} 
$(N-1)$ was approximated by $N$. 
The remaining cumulant in \eqref{G17} can be easily evaluated by use 
of Eq. \eqref{G7} and the matrix elements of Tab. \ref{eigen}. We find
\begin{eqnarray}
  {\cal H}_{\rm eff}(\beta\rightarrow\infty)|_{\rm 2nd\, order} &=&
  - \frac{3}{32} N \Delta \lambda^{2} (1-2\alpha)^{2} \, {\cal P} .
  \nonumber \\[-1ex]
  \label{G18}
  &&
\end{eqnarray}

Next, we use the computer to calculate the higher 
orders of perturbation theory.  By use of the computer algebra 
system Maple V \cite{Maple} we have implemented the following steps:
\begin{itemize}
  \item[(a)]
  At first, the decomposition \eqref{G9} of the perturbation ${\cal H}_{1}$ is 
  inserted in the cumulant expressions so that all energy denominators can 
  be easily evaluated [compare with Eq.~\eqref{G6}]. Consequently, the 
  cumulants of Eq.~\eqref{G4} are decomposed into sums of expressions of the 
  form \eqref{G14}.
  \item[(b)]
  Cumulants of the form \eqref{G14} are computed by use of Eq.~\eqref{G7} if 
  the above conditions (i) - (iii) are fulfilled. Otherwise no 
  further evaluation is needed. 
  \item[(c)]
  Finally, the remaining operator products of Eq.~\eqref{G7} are calculated.
\end{itemize}
It is obvious that steps (a) and (b) can easily be implemented by 
use of computer loops. 
For the calculation of the operator products of 
Eq.~\eqref{G7} in step (c), it is sufficient to  consider finite clusters. 
The finite order 
contribution of a short-ranged perturbation is independent from the cluster 
size if the cluster is chosen large enough. One can also 
prove explicitly that the results are not affected 
by the cluster size. A state of the 
cluster is described by an array of integers $\{1,2,3,4\}$ 
which represent the states $\{s,t^{+},t^{0},t^{-}\}$ for each 
dimer. The ${\cal T}$ operators applied to a cluster 
change the elements of the integer array which then describes 
the final cluster state. Finally, one has to count over 
all contributing processes.

As result of the computer aided perturbation theory we find 
the following effective Hamiltonian 
\begin{eqnarray}
  \label{G19}
  {\cal H}_{\rm eff}(\beta\rightarrow\infty) &=& 
  - N J \sum_{n} 
  \left(
    \frac{\lambda}{4}
  \right)^{n}
  h_{n}
  \, {\cal P}
\end{eqnarray}
The parameters $h_{n}$ up to seventh order read
\begin{eqnarray}
  h_{0} &=& \frac{3}{4}, \label{G20} \\[1ex]
  h_{1} &=& 0,  \nonumber \\[1ex]
  h_{2} &=& \frac{3}{2}\left(1-2\alpha\right)^{2}, \nonumber \\[1ex]
  h_{3} &=& 
    \left(\frac{3}{2}+3\alpha\right)\left(1-2\alpha\right)^{2}, 
    \nonumber \\[1ex]
  h_{4} &=& 
    \left(
      \frac{13}{8}+\frac{27}{2}\alpha-\frac{3}{2}\alpha^{2}
    \right)
    \left(1-2\alpha\right)^{2}, \nonumber \\[1ex]
  h_{5} &=& 
    \left(
      \frac{89}{24}+\frac{311}{12}\alpha+\frac{93}{2}\alpha^{2}-45\alpha^{3}
    \right)
    \left(1-2\alpha\right)^{2}, \nonumber \\[1ex]
  h_{6} &=& 
    \left(
      \frac{463}{48} + \frac{454}{9}\alpha + 
      \frac{1307}{6}\alpha^{2}-84\alpha^{3}-159\alpha^{4}
    \right) \times
    \nonumber \\
    && \quad \times \,
    \left(1-2\alpha\right)^{2}, \nonumber \\[1ex]
  h_{7} &=& 
    \left(
      \frac{81557}{3456} + \frac{257909}{1728}\alpha + 
      \frac{215995}{432}\alpha^{2} + \frac{173579}{216}\alpha^{3} 
    \right.
    \nonumber \\
  && \quad -\,
    \left. 
      \frac{14865}{8}\alpha^{4} + \frac{879}{4}\alpha^{5}
    \right)
    \left(1-2\alpha\right)^{2}. \nonumber
\end{eqnarray}
Apart from the projector ${\cal P}$ the effective Hamiltonian 
[\eqref{G19} and \eqref{G20}] can be directly identified as the 
ground-state energy of the original Hamiltonian \eqref{G1}. As discussed
above, this follows
from the fact that the low-energy subspace ${\cal U}_{P}$ only consists 
of a single state. Note that the present result for 
the ground-state energy agrees with the 
result obtained before on the basis of the flow equation method 
\cite{Knetter}.


\section{The generalized model} \label{general}

The unperturbed part ${\cal H}_{0}$ of the Hamiltonian \eqref{G1}, which we
have studied up to now, has had an equidistant spectrum. Note that this
property of ${\cal H}_{0}$ was explicitly needed in the flow-equation 
approach to construct perturbative effective Hamiltonians \cite{Knetter}. 
In the following we want to demonstrate that this
property of the model Hamiltonian is not required in the cumulant method 
discussed above. For that purpose let us modify the unperturbed Hamiltonian 
${\cal H}_{0}$ of the model \eqref{G1}, whereas ${\cal H}_1$
is kept unchanged: The coupling between the two 
spins of each dimer is chosen to be different for dimers with even and 
odd site index $j$
\begin{eqnarray}
  \label{G21}
  {\cal H}_{0} &=& 
  J \sum_{j=1}^{N}{\bf s}_{j}^{e}\cdot{\bf s}_{j}^{o} + 
  J \delta\sum_{j=1}^{N/2} {\bf s}_{2j}^{e}\cdot{\bf s}_{2j}^{o}.
\end{eqnarray}
Note that the new Hamiltonian ${\cal H}_0$ does not change the 
eigenstates of the previous unperturbed 
Hamiltonian \eqref{G1}. However, the  eigenenergies are different. 
The dimer singlet and triplet energies depend on the dimer index j
and are given by 
\begin{eqnarray}
  \label{G22}
  \varepsilon_{s}(j) &=& 
  \left\{ \begin{array}{*{2}{l}}
    \varepsilon_{s} &  \qquad \mbox{$j$ odd} \\ 
    (1+\delta) \, \varepsilon_{s} & \qquad \mbox{$j$ even}
  \end{array}\right. \\[2ex]
  \label{G23}
  \varepsilon_{t}(j) &=& 
  \left\{ \begin{array}{*{2}{l}}
      \varepsilon_{t} & \qquad \mbox{$j$ odd} \\ 
      (1+\delta) \, \varepsilon_{t} & \qquad \mbox{$j$ even}
  \end{array}\right. 
\end{eqnarray}
Therefore, the spectrum of the unperturbed Hamiltonian ${\cal H}_{0}$ 
is not equidistant anymore. 
Due to the modification \eqref{G21} of ${\cal H}_0$ also the 
Liouville operator ${\bf L}_0$ has changed.
The decomposition of
the perturbation ${\cal H}_{1}$, Eq.~\eqref{G1},
into transition operators with respect to ${\cal H}_0$  has to be
modified too. 
In particular, the number of created local triplets can not be the only
classification criterion anymore since the singlet-triplet splittings
differs for different dimer sites. 
In fact, the former transition operators
${\cal T}_{1}$, ${\cal T}_{0}$, and ${\cal T}_{-1}$ have to be split up. 
The perturbation ${\cal H}_{1}$ can now be written as
\begin{eqnarray}
  \label{G24}
  {\cal H}_{1} &=& {\cal T}_{-2} + {\cal T}_{-1,-} +{\cal T}_{-1,0} + 
  {\cal T}_{0,-} + {\cal T}_{0,0} \\
  && 
  + \, 
  {\cal T}_{0,+} + {\cal T}_{1,0} + 
  {\cal T}_{1,+} + {\cal T}_{2} \nonumber
\end{eqnarray}
and the respective eigenvalues of the modified Liouville operator now
read
\begin{eqnarray}
  \label{G25}
  \Delta_{2} &=& (2 + \delta) \, \Delta\\
  \nonumber
  \Delta_{1,+} &=& (1 + \delta) \, \Delta\\
  \nonumber
  \Delta_{1,0} &=& \Delta \\
  \nonumber
  \Delta_{0,+} &=& \delta \, \Delta \\
  \nonumber
  \Delta_{0,0} &=& 0\\
  \nonumber
  \Delta_{0,-} &=& -\delta \, \Delta \\
  \nonumber
  \Delta_{-1,0} &=& -\Delta \\
  \nonumber
  \Delta_{-1,-} &=& -(1+\delta) \, \Delta \\
  \nonumber
  \Delta_{-2} &=& -(2 + \delta) \, \Delta
\end{eqnarray}
The second index $x$ in ${\cal T}_{\pm1,x}$ and  ${\cal T}_{0,x}$ 
denotes how  the number of triplets on dimers with even index $j$ is changed. 
($x=+(-)$ describes  the creation (annhiliation) of a triplet and  
$x=0$ no change). Note that the
original model \eqref{G1} is given by $\delta=0$.

The next steps can be done as before. At first we introduce local 
transition operators. Furthermore, the restrictive conditions (a), (b), (c) 
are still valid so that the evaluation of the cumulants can be done similar
as before. (Of course, now one has to distinguish  
between odd and even dimer indices.)
By use of the computer aided perturbation theory we now find
an effective Hamiltonian which has the form of expression \eqref{G19}.
The parameters $h_{n}$, up to fourth order,  now read
\begin{eqnarray}
  \label{G26}
  h_{0} &=& \frac{3}{8}(2+\delta), \\[1ex]
  \nonumber
  h_{1} &=& 0, \\[1ex]
  \nonumber
  h_{2} &=& \frac{3}{(2+\delta)}\left(1-2\alpha\right)^{2}, \\[1ex]
  \nonumber
  h_{3} &=& \frac{4}{(2+\delta)^{2}} 
  \left(\frac{3}{2}+3\alpha\right)\left(1-2\alpha\right)^{2}, \nonumber \\[1ex]
  h_{4} &=& 3 \frac{(1-2\alpha)^2}{(2+\delta)^2} 
  \left(
    12 \frac{2+\delta}{(3+2\delta)(3+\delta)} 
  \right.
  \nonumber \\
  && 
  \left.
    \quad +\,
    \frac{(2+\delta)(1-2\delta)^2}{1+\delta} -
    \frac{5-52 \alpha + 20\alpha^2}{2+\delta} 
  \right) .
  \nonumber
\end{eqnarray}
In fact, we have calculated the $h_{n}$ also up to seventh order.
Since the expressions are rather involved, here the $h_n$'s are only given   
up to the fourth order. Higher orders are 
available on request. Note that in the case of $\delta=0$ Eq.~\eqref{G26} 
reduces to \eqref{G20}.


\section{Conclusions} \label{Conc}

In this paper we have shown that a recently developed cumulant 
method \cite{Hubsch} can be used to develop computer aided perturbation 
theory. Size consistency of extensive variables is fulfilled. The 
cumulant method offers compact expressions for the different orders of the 
perturbation theory. Furthermore, the evaluation of the cumulant expressions 
is reduced to the problem how to properly count the contributing processes. 

We have applied the cumulant method to the dimerized and frustrated spin 1/2 
chain. For this model the ground-state energy was calculated up to 
seventh order perturbation theory where our results agree with those 
obtained by the flow equation method \cite{Knetter}. 
It turned out that the efficiency 
of the computer based evaluation of the cumulant expressions can be 
enormously improved if restrictive conditions are considered.  In 
this way the number of vanishing contributions in the calculations 
can be reduced. Furthermore, we have modified the model by
an additional site-oscillating dimer coupling so that the spectrum of the
unperturbed part of the Hamiltonian is not equidistant anymore. For the
generalized model we have calculated the ground-state energy up to seventh
order perturbation theory as well. Note that in contrast to the cumulant
approach the flow equation method used in Ref.~\onlinecite{Knetter} requires
an equidistant spectrum of the unperturbed 
part of the Hamiltonian. On the other hand, the flow equation 
method can also be used to calculate excitation energies.

The derivation of the cumulant expression
\eqref{G4} for the effective Hamiltonian 
${\cal H}_{\mbox{eff}}(\beta \rightarrow \infty)$
was  based on the assumption that it is acting in the 
lowest energy subspace ${\cal U}_P$  of ${\cal H}_0$. This subspace 
can either be degenerate or nondegenerate. 
For the dimerized spin $1/2$ chain the
lowest energy subspace is one-dimensional and  is 
given by the product of all dimer singlet states. 
An example for a degenerate unperturbed groundstate 
was discussed in Ref.~\onlinecite{Hubsch}.  Note that 
the cumulant expression \eqref{G4} is also 
closely related to an effective Hamiltonian which was 
derived by Takahashi \cite{Takahashi}. There,  
the eigenvalue problem of the full Hamiltonian is transformed 
to that of an effective Hamiltonian which acts in a degenerate 
or nondegenerate energy subspace 
of ${\cal H}_0$. By comparing the perturbation expansion \eqref{G4}
order by order with that of Ref.~\cite{Takahashi} the equivalence 
of both approaches can be shown. Takahashi's approach  
does not involve the  temperature. Therefore, the subspace in which 
the effective Hamiltonian acts is not necessarily the 
lowest energy subspace of ${\cal H}_0$. 
Thus, one might expect the cumulant result \eqref{G4} 
should also be valid for this case 
which would allow to calculate excitation energies. Finally,    
a projector-based renormalization method (PRM) for 
effective Hamiltonians  was recently introduced by 
two of the present authors \cite{SommerHubschBecker}. 
By using perturbation theory also in this approach 
\cite{tobepublished} a close relation to the cumulant 
expression \eqref{G4} can be found. 
However, the PRM treatment seems to be more suited to calculate excitation
energies  than the cumulant approach presented here.

\section*{Acknowledgments}
We would like to acknowledge helpful discussions with K.~Meyer, T.~Sommer, 
and P.~Zahn. This work was supported by the DFG through the research program 
SFB 463 and under Grant No. HU~993/1-1.



\begin{thebibliography}{}

\bibitem{Schrieffer} J.R. Schrieffer and  P.A. Wolff, Phys. Rev. ${\bf 149}$,
  491 (1966).
\bibitem{Fulde} See, for example, P. Fulde, 
  \emph{Electron Correlations in Molecules and Solids} 
  (Springer-Verlag, Berlin, 3. ed. 1995), Chap. 4.2.
\bibitem{Wegner} F. Wegner, Ann. Phys. (Leipzig) {\bf 3},77 (1994).
\bibitem{Glazek} S.D.~G{\l}azek and K.G.~Wilson, Phys. Rev. D {\bf 48}, 5863
  (1993); S.D.~G{\l}azek and K.G.~Wilson, Phys. Rev. D {\bf 49}, 4214 (1994).
\bibitem{Stein} J. Stein, J. Stat. Phys. {\bf 88}, 487 (1997). 
\bibitem{Takahashi} M. Takahashi, J. Phys. C: Solid State Phys. {\bf 10}, 
  1289 (1977).
\bibitem{M-Hartmann} E. M\"{u}ller-Hartmann and A. Reischl, Eur. Phys. J 
  B {\bf 28}, 173 (2002).
\bibitem{Knetter} C. Knetter and G.S. Uhrig, Eur. Phys. J B {\bf 13}, 209 
  (2000).
\bibitem{Hubsch} A. H\"{u}bsch, M. Vojta and K.W. Becker, J. Phys: Condensed
  Matter ${\bf 11}$, 8523 (1999).
\bibitem{Riera} J. Riera and A. Dobry, Phys. Rev. B {\bf 51}, 16098 (1995).
\bibitem{Castilla} G. Castilla, S. Chakravarty, and V. J. Emery, Phys. Rev. 
  Lett. {\bf 75}, 1823 (1995).
\bibitem{Bray} J.W. Bray, L.V. Interante, I.C. Jacobs, J.C. Bonner, in 
  \emph{
    Extended Linear Chain Compounds, 
  } edited by J.S. Miller (Plenum Press, New York, 1983), Vol. 3, p. 353.
\bibitem{Polatsek} G. Polatsek and K.W. Becker, Phys. Rev. B {\bf 55}, 16096 
  (1997).
\bibitem{Kubo} R. Kubo, J. Phys. Soc. Jpn. {\bf 17}, 1100 (1962).
\bibitem{Kladko} K. Kladko and P. Fulde, Int. J. Quantum. Chem. {\bf 66}, 377 
  (1998).
\bibitem{Maple} See, for example, M. Abell and J. Braselton, 
  \emph{The Maple V Handbook} (Academic Press, San Diego, 1994).
\bibitem{SommerHubschBecker} K. W. Becker, A.~H\"{u}bsch, and T. Sommer, 
  Phys. Rev. B {\bf 66}, 235115 (2002). 
\bibitem{tobepublished} K.W.~Becker, A.~H\"ubsch, to be published. 
\end{thebibliography}
\end{document}